\documentclass[prd,aps,showpacs,preprintnumbers,amsmath, amssymb]{revtex4-2}

\usepackage{graphics,epsfig,hyperref}
\usepackage{graphicx}
\usepackage{dcolumn}
\usepackage{bm}
\usepackage{epstopdf}
\usepackage{color}
\usepackage{xcolor}
\usepackage{subfigure}
\usepackage{diagbox} 
\usepackage{adjustbox}
\usepackage{multirow}
\usepackage{amsmath,amssymb,amsfonts}

\usepackage[normalem]{ulem}
\usepackage{xcolor}
\usepackage{mathtools} 
\usepackage{amsthm,amssymb}
\usepackage{mathrsfs}

\begin{document}
	\title{\boldmath Stationary scalar clouds around a rotating Kalb-Ramond BTZ black hole}

	\author{Rui Ding$^{1}$, Fangli Quan$^{1}$, Zhong-Wu Xia$^{1}$\footnote{zwxia@hunnu.edu.cn}, Qiyuan Pan$^{1,2}$\footnote{panqiyuan@hunnu.edu.cn} and Jiliang Jing$^{1,2}$\footnote{jljing@hunnu.edu.cn}}
	\affiliation{	$^1$Department of Physics, Institute of Interdisciplinary Studies, Key Laboratory of Low Dimensional Quantum Structures and Quantum Control of Ministry of Education, Synergetic Innovation Center for Quantum Effects and Applications, and Hunan Research Center of the Basic Discipline for Quantum Effects and Quantum Technologies, Hunan Normal University,  Changsha, Hunan 410081, People's Republic of China} 
	\affiliation{$^2$Center for Gravitation and Cosmology, College of Physical Science and Technology, Yangzhou University, Yangzhou 225009, People's Republic of China
	}
	
	\begin{abstract}
		\baselineskip=0.6 cm
	\begin{center}
			{\bf Abstract}
	\end{center}		
	We investigate the scalar clouds around a rotating Kalb-Ramond (KR) BTZ black hole under Robin boundary conditions. The clouds are obtained as stationary bound states at the superradiant threshold $\omega=m\Omega_H$, where the KR parameter, the rotation and the Robin boundary jointly determine
	their existence. It is shown that the KR parameter qualitatively changes the existence lines of
	clouds. For a nonpositive KR parameter, the lines remain monotonic, whereas for a positive KR
	parameter they can become nonmonotonic, so that a fixed boundary condition may admit clouds
	in disconnected regions of parameter space. Quasinormal modes (QNMs) and horizon fluxes are
	further used as consistency checks, confirming that the cloud solutions correspond to non-damping
	modes at the superradiant threshold where the energy flux changes sign. The KR parameter also
	shifts the critical Robin parameter at which the clouds exist. These results establish stationary scalar
	clouds as sensitive probes of the interplay between the Robin boundary conditions and KR gravity.
\end{abstract}

	\maketitle
	\flushbottom

	\newcommand{\Heun}{\text{$H_G$}}
	
\section{Introduction}
To resolve the singularity problem inherent in classical general relativity, quantum gravity has long been regarded as one of the central directions in fundamental physics~\cite{Birrell:1982ix,Maldacena:1997re,Aharony:1999ti,Gubser:1998bc,Alfaro:1999wd,Alfaro:2001rb,Rovelli:1989za}. Some approaches to the quantum gravity suggest that the Lorentz symmetry may be broken at large distance and low-energy scales, where the gravitational phenomena are directly accessible to current observations~\cite{Liu:2025fxj}. Therefore, the Lorentz symmetry breaking (LSB) gravity has attracted considerable attention in recent years~\cite{Lessa:2025kln,Zhang:2025kcw,Zeng:2024ptv,Xia:2025yzg,Li:2025tcd,Do:2020ojg,Liu:2024oas,Duan:2023gng,Yang:2023wtu,Lessa:2019bgi,Kumar:2020hgm,Liu:2024lve,Liu:2025oho,Liu:2025bpp}. Among the various realizations of LSB, the Kalb-Ramond (KR) field provides a particularly well-motivated framework, since it naturally emerges as a rank-2 antisymmetric tensor field in string theory and can trigger the spontaneous LSB once it acquires a nonzero vacuum expectation value (VEV)~\cite{Kalb:1974yc,Manton:2024hyc,Capanelli:2023uwv,Shi:2025rfq,Malta:2025ydq}. Within the KR-gravity framework, exact static and spherically symmetric black-hole solutions have been constructed, providing KR-deformed generalizations of the Schwarzschild-(A)dS geometry and a basis for studying phenomenological signatures of the antisymmetric tensor background~\cite{Yang:2023wtu,Sucu:2025lqa}. In lower-dimensional settings, exact  dilaton black hole and black string solutions supported by the KR field have also been obtained~\cite{Asrat:2024cug,Cordeiro:2025cfo}.
 More recently, a rotating KR BTZ black hole solution was  constructed by Liu et al.~\cite{Liu:2025fxj}. This exact asymptotically AdS solution contains a continuously tunable KR parameter $\ell$ and reduces to the standard rotating BTZ black hole in the limit $\ell\to0$. Since the KR background effectively deforms the cosmological constant and the horizon structure while preserving an analytic metric form, it provides a useful lower-dimensional test for probing how the	 spontaneous LSB influences black hole physics. 

The aforementioned works make the KR BTZ black hole an appealing background for studying matter field configurations in LSB spacetimes.  In particular, our main interest lies in scalar clouds, namely stationary bound states that arise when the scalar field frequency is synchronized with the angular velocity of the horizon~\cite{Herdeiro:2014goa,Herdeiro:2015gia,Hod:2013zza,Hod:2014baa,Benone:2014ssa,Herdeiro:2014pka,Hod:2014sha,Hod:2015goa,Hod:2015bdw,Hod:2015wwy,Hod:2016lgi}. This synchronization condition precisely marks the threshold of superradiance, so scalar clouds provide a particularly clean probe of how the KR deformation influences the onset of superradiance in black hole spacetimes. The scalar clouds are also of particular interest because, if sufficiently populated through the superradiant amplification, the scalar cloud can backreact on the background geometry and eventually lead to a hairy black hole solution~\cite{Herdeiro:2014goa,Ferreira:2017cta}. In this context, it is natural to ask whether the KR BTZ black hole can also support scalar clouds under vanishing energy flux boundary conditions~\cite{Ferreira:2017cta}, and, if so, how the KR parameter $\ell$ affects their  existence lines and radial structure. This naturally motivates our systematic study of scalar clouds in the KR BTZ background. 
	
In order to properly understand the cloud solutions, it is also useful to examine the associated quasinormal mode (QNM) spectrum and the role of superradiance~\cite{Lutfuoglu:2025kqp,Konoplya:2024lch}. QNMs encode the dynamical response of the spacetime to scalar perturbations and can therefore provide complementary information on how the KR parameter affects the behavior of the system near the cloud regime. Likewise, since scalar clouds arise exactly at the onset of superradiance, clarifying the superradiant condition is essential for identifying and interpreting these stationary states. At the same time, the situation in asymptotically AdS spacetimes is more subtle: modes that grow in time do not necessarily correspond to the energy extraction from the black hole, but may instead be driven by the bulk instability~\cite{Garcia:2018sjh,Casana:2017jkc}. In this sense, not every instability should be identified as superradiance. Therefore, although our primary focus is on scalar clouds, the analysis of QNMs and superradiance plays an important supporting role, helping to clarify the dynamical origin, existence conditions, and physical interpretation of the cloud configurations in the KR BTZ spacetime.

Based on above considerations, this work aims to provide a systematic investigation of the existence of
scalar clouds around a rotating Kalb-Ramond BTZ black hole by separating the scalar perturbation equation
and mapping the resulting radial equation into a general Heun form~\cite{Bertolami:2005bh}. 
For scalar clouds, we first impose
the synchronization condition $\omega=m\Omega_H$, and then require the ingoing behavior at the horizon together with
vanishing energy flux boundary conditions at infinity. The resulting radial equation is analyzed through the
Wronskian construction~\cite{Bailey:2006fd,Kinoshita:2023iad,Wu:2025wbp}, from which we extract the existence lines of the stationary cloud configurations. For QNMs and superradiance, by contrast, the frequency is treated as an eigenvalue determined directly
by the same boundary conditions. Here we examine how the KR parameter $\ell$ and the azimuthal number $m$
affect the cloud existence lines, and the connection between scalar cloud with QNMs and superradiance.

Our work is organized as follows. In Section II, we briefly review the rotating KR BTZ black hole geometry, and then introduce the massive Klein-Gordon equation together with the appropriate boundary conditions. In Section III, we analyze stationary scalar configurations around the rotating KR BTZ black hole, and present the results of QNMs and superradiance.  Finally, in the last section we reach our conclusions.

\section{Background geometry, massive scalar fields and boundary conditions}\label{Sec.II}

\subsection{Rotating KR BTZ  black holes}
We begin with the Einstein-Hilbert action nonminimally coupled to a rank--2 antisymmetric KR tensor $B_{\mu\nu}$ in the form~\cite{Altschul:2009ae}
	\begin{equation}
		\label{eq:action}
		S=\int d^D x \,\sqrt{-g}\,
		\left[
		\frac{1}{2\kappa}\bigl(R-2\Lambda\bigr)
		-\frac{1}{12} H_{\mu\nu\rho}H^{\mu\nu\rho}
		- V\!\bigl(B_{\mu\nu}B^{\mu\nu}\pm b^2\bigr)
		+\frac{1}{2\kappa}\Bigl(\xi_1 B_{\mu\nu}B^{\mu\nu} R
		+ \xi_2 B_{\rho\mu} B^{\ \mu}_{\nu} R^{\rho\nu}\Bigr)
		\right],
	\end{equation}
where \(\Lambda\) denotes the cosmological constant and \(\kappa\) represents the gravitational coupling constant. The coupling constants \(\xi_1\) and \(\xi_2\) govern the interaction between gravity and the KR field. The KR field strength $H_{\mu\nu\rho}\equiv\partial_{[\mu}B_{\nu\rho]}$. The potential \(V\) is chosen to ensure a nonzero VEV for the KR field. It should be noted that the nonzero VEV selects a specific direction, which results in the violation of local Lorentz invariance~\cite{Ding:2023niy,Liu:2024oas}.

Varying the action \eqref{eq:action}, we can obtain the stationary, axially symmetric black hole metric in the three dimensional spacetime ($D=3$)~\cite{Liu:2025fxj}
	
\begin{equation}
	\mathrm{d} s^{2} = - A(\tilde{r}) \mathrm{d} t^{2} + S(\tilde{r}) \mathrm{d} \tilde{r}^{2} + \tilde{r}^{2} [K(\tilde{r}) \mathrm{d} t + \mathrm{d} \varphi]^{2} , \label{eq:metric}
\end{equation}
with
\begin{align}
	A(\tilde{r}) = & - M - \frac{\Lambda}{(1 + \ell) } \tilde{r}^{2} + \frac{j^{2}}{4 (1 + \ell) \tilde{r}^{2}} , \notag\\
	S(\tilde{r}) = & \frac{1}{A(\tilde{r})} , \notag\\
	K(\tilde{r}) = & - \frac{j}{2 \tilde{r}^{2}} ,
\end{align}
where $\Lambda = -1/\lambda^2$ is the cosmological constant with the AdS radius $\lambda$, $\ell$ is the KR parameter, and 
$M$ and $j$ are the mass and the spin parameters of the black hole. Obviously, this solution reduces to the rotating BTZ black hole~\cite{Banados:1992wn} when $\ell$= 0. For simplicity, we introduce a new radial coordinate $r$ via the transformation
\begin{equation}
\tilde{r} \rightarrow r = \tilde{r}^{2}.
\end{equation}
Thus, the black hole solution \eqref{eq:metric} has two horizons: an inner horizon at $r_{-}$ and an outer horizon at $r_{+}$, both of which depend on the KR parameter $\ell$
	\begin{equation}
		r_{\pm} = \frac{1}{2} \lambda \left[\lambda(1+\ell) M\pm\sqrt{\lambda ^2 (1+\ell)^2 M^2-j^2}\right].
	\end{equation}
Furthermore, the angular velocity of the horizon is given by
\begin{equation}\label{angularmom}
	\Omega_H =\frac{\sqrt{r_-} }{\lambda\sqrt{ r_+}}.
\end{equation}
Therefore the mass $M$ and angular momentum $j$ can be expressed as
\begin{equation}\label{exeq}
	M = \frac{r_+ + r_-}{\lambda ^2 \left (1+\ell   \right ) } = \frac{r_+\left( 1 + \lambda ^2 \Omega^2_H \right)}{\lambda ^2 \left ( 1+\ell  \right ) } , \quad j=2\frac{\sqrt{r_{+}r_{-}}}{\lambda}=2r_+\Omega_H.
\end{equation}
For the extremal KR BTZ black hole, we have $r_+ =r_- = \lambda^{2} (1+\ell) M/2$, and the angular velocity of the horizon is completely determined by the AdS radius $\Omega_H =1/\lambda$. In this case, the black hole mass is related to $\Omega_H$ by $M=2r_{+}\Omega_H^{2}/\left ( 1+\ell  \right ) $.

\subsection{Massive Klein-Gordon equation}	
In the rotating KR BTZ black hole background \eqref{eq:metric}, we consider a massive scalar field $\Phi$, with the effective squared mass $\mu^2/\lambda^2$, which satisfies the Klein-Gordon equation \cite{Dappiaggi:2017pbe}
	\begin{align}
		\left( \nabla^\mu\nabla_\mu - \frac{\mu^{2}}{\lambda^2} \right) \Phi(t,r,\varphi) = 0,
	\end{align}
and separate the variables of the scalar field
    \begin{align}
        \Phi(t,r,\varphi) = e^{- i \omega t + i m \varphi} \phi(r),
    \end{align}
where $\omega$ and $m$ are the frequency and the azimuthal quantum number of the scalar field. Then we obtain the radial equation 	
	\begin{align}
		\nonumber & \left\{ \frac{\mathrm{d}^{2} }{\mathrm{d} r^{2}} + \left( \frac{1}{r - r_{+}} + \frac{1}{r - r_{-}} \right) \frac{\mathrm{d}}{\mathrm{d} r} + \frac{\lambda^{4}(1 + \ell)^{2}}{16 r (r - r_{+})^{2}(r - r_{-})^{2}} \right. \\
		& \left. \times\left[(2 \omega r - m j) ^{2} - \frac{4 (r - r_{-}) (r - r_{+}) (\mu^{2} r + m^{2}\lambda^2)}{\lambda^{4} (1 + \ell)} \right] \right\} \phi(r) = 0 . \label{eq:RadialEqu}
	\end{align}
To solve this equation, we introduce another radial coordinate $z = \frac{r - r_{+}}{r - r_{-}}$. This transformation establishes a point-to-point mapping
	\begin{equation}
		f: r \rightarrow z \Rightarrow (r_{+}, \infty, 0, r_{-}) \rightarrow (0, 1, a, \infty),
	\end{equation}
where we have defined $a=r_{+}/r_{-}$. So the radial equation~\eqref{eq:RadialEqu} can be rewritten as
	\begin{equation}
		z (1 - z) {\phi}''(z) + (1 - z) {\phi}'(z) + \left( \frac{A_{0}}{z} + \frac{A_{1}}{z - 1} + \frac{A_{a}}{z - a} + B \right) \phi(z) = 0 ,\label{eq:Heun}
	\end{equation}
with
	\begin{align}
		A_{0} = \frac{(2 \omega r_{+} - m j)^{2} (1 + \ell)^{2}}{16 r_{+} (r_{+} - r_{-})^{2}} \lambda^{4} ,\quad
		A_{1} = \frac{(1 + \ell) \mu^{2}}{4} ,\quad\notag\\
		A_{a} = - \frac{m^{2} j^{2} \ell (1 + \ell)}{16 r_{+} r_{-}^{2}}\lambda^{4} , \quad
		B =  - \frac{(2 \omega r_{-} - m j)^{2} (1 + \ell)^{2}}{16 r_{-} (r_{+} - r_{-})^{2} }\lambda^{4} .\label{eq:A01aB}
	\end{align}
It is easy to check that when the KR parameter $\ell=0$, the radial equation \eqref{eq:Heun} reduces to the BTZ black hole case with $A_a=0$, i.e., the standard Gauss hypergeometric equation. The radial equation \eqref{eq:Heun} we obtain has four regular singular points, \((0,1,a,\infty)\), and therefore belongs to the Fuchsian class of differential equations~\cite{olver2010nist}. In principle, it can be transformed into the general Heun equation. Therefore, we use the following redefinition of the radial function
\begin{align}
\phi(z) \rightarrow z^{\alpha_{1}} (z - 1)^{\beta_{1}} (z - a)^{\delta_{1}} \tilde{\phi}(z),
\end{align}
and transform Eq.~\eqref{eq:Heun} into
\begin{equation}
{\tilde{\phi}}''(z) + \left( \frac{1 + 2 \alpha_{1}}{z} + \frac{2 \beta_{1}}{z - 1} + \frac{2 \delta_{1}}{z - a} \right) {\tilde{\phi}}'(z) + \sum_{k = 1}^{2} \left[\frac{\Theta_{k}}{z^{k}} + \frac{\Psi_{k}}{(z - 1)^{k}} + \frac{\Delta_{k}}{(z - a)^{k}} \right] \tilde{\phi}(z) = 0 , \label{eq:NewRadialEqu}
\end{equation}
with
	\begin{align}
		\Theta_{1} = & -\frac{a (-A_{0}+2 \alpha_{1} \beta_{1}+A_{1}-B+\beta_{1})+2 \alpha_{1} \delta_{1}+A_{a}+\delta_{1}}{a} , \notag\\
		\Psi_{1} = & \frac{(a-1) (A_{1} + 2 \alpha_{1} \beta_{1} - B + \beta_{1} -A_{0}) + A_{a} - 2 \beta_{1} \delta_{1}}{a-1} , \notag\\
		\Delta_{1} = & \frac{(a-1) (2 \alpha_{1}+1) \delta_{1}+2 a \beta_{1} \delta_{1}-A_{a}}{(a-1) a}, \notag\\
		\Theta_{2} = & A_{0}+\alpha_{1}^2 , \notag\\
		\Psi_{2} = & -A_{1}+\beta_{1}^2-\beta_{1} , \notag\\
		\Delta_{2} = & \delta_{1}^2-\delta_{1}.
	\end{align}
However, the standard form of the Heun equation is~\cite{Hortacsu:2011rr}
\begin{equation} \label{eq:Heun-Equation}
	{\tilde{\phi}}''(z) + \left( \frac{\gamma}{z} + \frac{\delta}{z - 1} + \frac{\epsilon}{z - a} \right) {\tilde{\phi}}'(z) + \frac{\alpha \beta z - q}{z (z - 1) (z - a)} \tilde{\phi}(z) = 0 .
\end{equation}
To transform Eq.\eqref{eq:NewRadialEqu} into the Heun equation, the following conditions must be satisfied
\begin{equation}	
	\gamma = 1 + 2 \alpha_{1},
	\delta = 2 \beta_{1},
	\epsilon = 2 \delta_{1},
	\Theta_{1} = -\frac{q}{a},
	\Psi_{1} = \frac{q-\alpha \beta }{a-1},
	\Delta_{1} = \frac{a \alpha \beta-q}{a\left (a-1\right ) }, 
	\Theta_{2} = 0, 
	\Psi_{2} = 0,
	\Delta_{2} = 0,
\end{equation}
which leads to
\begin{equation}	
		\gamma = 1 + 2 \alpha_{1},
		\delta = 2 \beta_{1},
		\epsilon = 0,
		\delta_{1} = 0, 
		\alpha_{1} = - i \sqrt{A_{0}},
		\beta_{1} = \frac{1}{2} \left (1-\sqrt{1+4 A_{1}}\right ), \label{eq:GDEDAB}
\end{equation}
and
\begin{equation}	
	\begin{cases}
		\alpha = \alpha_{1}+\beta_{1}+\sqrt{B}, \\
		\beta = \alpha_{1}+\beta_{1}-\sqrt{B}, \\
		q = \alpha \beta a + A_{a}. \label{eq:ABq}
	\end{cases}
\end{equation}
Thus, the corresponding solution at $z=0$ is the Heun function $\Heun(a,q;\alpha, \beta, \gamma, \delta; z)$.

\subsection{Boundary conditions}

Similar to searching for QNMs, finding scalar clouds also involves solving an eigenvalue problem for a differential equation. When solving for QNMs, we take the scalar field frequency $\omega$ to be complex. However, to obtain stationary scalar clouds, we instead assume $\omega$ is real, which implies that the field neither grows nor decays in time. For both QNMs and scalar clouds, we impose the same physical conditions: an ingoing wave at the event horizon and the vanishing energy flux, which corresponds to a Robin boundary condition ~\cite{Ferreira:2017cta,Dappiaggi:2017pbe,Wang:2016zci,Wang:2015goa}, at spatial infinity. The solutions to Eq. \eqref{eq:RadialEqu} at event horizon ($z=0$) can be expressed in terms of the Heun function
	\begin{align}
		\phi^{(\mathcal I)}(z) = & h(r) \Heun(a,q;\alpha, \beta, \gamma, \delta; z), \label{eq:I}\\
		\phi^{(\mathcal O)}(z) = & h(r) z^{1 - \gamma} \Heun(a,(a \delta + \epsilon)(1 - \gamma) + q;\alpha + 1 - \gamma, \beta + 1 - \gamma, 2 - \gamma, \delta; z), 
	\end{align}
with $h(r)={(r - r_{+})^{\alpha_{1}} (r_{-} - r_{+})^{\beta_{1}} (1 - a)^{\delta_{1}} r^{\delta_{1}}}/{(r - r_{-})^{\alpha_{1} + \beta_{1} + \delta_{1}}}$. Here, \(\mathcal I\) and \(\mathcal O\) represent the ingoing wave and outgoing wave at the horizon $z=0$, respectively. Consequently, the general solution of Eq. \eqref{eq:RadialEqu} at the horizon can be given by a linear combination of $\phi^{(\mathcal I)}(z)$ and $\phi^{(\mathcal O)}(z)$, $i.e.$,
\begin{equation}
	\phi(z)= A \phi^{(\mathcal I)}(z)+B \phi^{(\mathcal O)}(z),
\end{equation}
where $A$ and $B$ are the constants of integration. Obviously, an ingoing wave boundary condition implies $B=0$. At infinity ($z=1$), the solutions to Eq. \eqref{eq:RadialEqu} are
	\begin{align}
		\phi^{(\mathcal N)}(z) = & h(r) \Heun(1- a, \alpha \beta - q; \alpha, \beta, \delta, \gamma; 1 - z), \label{eq:neu}\\ 
		\phi^{(\mathcal D)}(z) = & h(r) (1 - z)^{1 - \delta}  \Heun(1 - a, ((1-a) \gamma + \epsilon)(1 - \delta) + \alpha \beta - q;\alpha + 1 - \delta, \beta + 1 - \delta, 2 - \delta, \gamma; 1 - z) ,\label{eq:Diri}
	\end{align}
where \(\mathcal{N}\) and \(\mathcal{D}\) denote the Neumann boundary condition and the Dirichlet boundary condition at infinity, respectively. Accordingly, the general solution at infinity is
\begin{equation}
		\phi(z) =C\phi^{(\mathcal N)}(z)+D\phi^{(\mathcal D)}(z),
\end{equation}
with two complex constants $C$ and $D$. Requiring the flux to vanish at infinity~\cite{Ferreira:2017cta,Wang:2021upj,Wang:2021uix} leads to the following Robin boundary condition~\cite{Ferreira:2017cta}
\begin{equation}\label{eq:robin}
	\phi^{(\mathcal R)}(z)=\sin(\xi)\phi^{(\mathcal N)}(z)+\cos(\xi) \phi^{(\mathcal D)}(z),~~\xi\in\left[0,\pi\right).
\end{equation}
Here, $\xi=0$ and $\xi=\pi/2$ correspond to the Dirichlet and Neumann boundary conditions, respectively.

Solving QNMs and scalar clouds can both be reduced to an eigenvalue problem for the radial equation. Given suitable boundary conditions, the corresponding eigenvalue spectrum can be determined. Specifically, we impose a purely ingoing boundary condition at the event horizon, while at infinity the solution is required to satisfy the condition of vanishing outgoing energy flux, which corresponds to a Robin boundary condition. Together with the appropriate matching condition, or equivalently the vanishing of the Wronskian determinant
\begin{equation}\label{eq:wronskian_def}
	W(z)=\phi^{(\mathcal R)}(z)\frac{\mathrm d}{\mathrm d z}\phi^{(\mathcal I)}(z)-\phi^{(\mathcal I)}(z)\frac{\mathrm d}{\mathrm d z}\phi^{(\mathcal R)}(z)=0,
\end{equation}
we can obtain the desired eigenvalues. In the following calculation, we set $z = 1/2$.
	
\section{Result}

\subsection{Stationary scalar clouds}
In this subsection, we investigate stationary scalar clouds around the rotating KR BTZ black hole. Scalar clouds are stationary bound states at the threshold of superradiance. They are characterized by a real frequency, and neither grow nor decay in time. This requirement implies that the frequency $\omega_c$ of the scalar cloud must be synchronized with the black hole horizon angular velocity, i.e.,
\begin{equation}\label{resonance}
	\omega=\omega_c = m \Omega_H.
\end{equation}
From \eqref{eq:A01aB}, we obtain
\begin{align}
	A_{0} = 0 ,\quad
	A_{1} = \frac{(1 + \ell) \mu^{2}}{4} ,\quad
	A_{a} = - \frac{m^{2} \ell (1 + \ell)}{4 r_{-}}\lambda^{2} ,\quad
	B = - \frac{m^{2} (1 + \ell)^{2}}{4 r_{+}}\lambda^{2}.
\end{align}
The coefficients in \eqref{eq:GDEDAB} and \eqref{eq:ABq} are now given by
\begin{equation}	
	\gamma = 1,\quad
	\delta = 2 \beta_{1},\quad
	\epsilon = 0,\quad
	\delta_{1} = 0, \quad
	\alpha_{1} = 0,\quad
	\beta_{1} = \frac{1}{2} \left (1 - \sqrt{1 + (1 +\ell) \mu^{2}} \right),
\end{equation}
and
\begin{equation}	
	\begin{cases}
		\alpha = \beta_{1} +i \frac{m \lambda}{2 \sqrt{r_{+} } } (1+\ell), \\
		\beta = \beta_{1} -i \frac{m \lambda}{2 \sqrt{r_{+} } } (1+\ell), \\
		q = a\beta_{1} ^{2}  +\frac{m^{2}\lambda^{2} }{4r_{-}} (1+\ell).
	\end{cases}
\end{equation}
Thus, the Eq.\eqref{eq:RadialEqu} has two branches of solutions at event horizon ($z=0$), one is a regular solution, which has a polynomial expansion near $z=0$, and the other is a logarithmically divergent solution. Based on $regularity$~\cite{Ferreira:2017cta}, we only take the regular solution
	\begin{align}
		\phi(z) = h(r) H_{G}\left (a,q;\alpha ,\alpha ^{\ast},1,2\beta _{1};z\right), 
	\end{align}
At infinity ($z=1$), Eqs.~\eqref{eq:neu}-\eqref{eq:Diri} reduce to
	\begin{align}
		\phi^{(\mathcal N)}(z) = &h(r) H_{G}\left (1-a,\alpha \alpha ^{\ast}-q ;\alpha ,\alpha ^{\ast},2\beta _{1},1;1-z\right) , \notag\\ 
		\phi^{(\mathcal D)}(z) = &h(r) \left(1-z\right)^{1-2\beta_{1}}H_{G}\left (1-a,\left (1-a\right)\left ( 1-2\beta _{1}\right) + \alpha \alpha ^{\ast}-q;\alpha-2\beta_{1}+1,\alpha ^{\ast}-2\beta_{1}+1,2-2\beta _{1},1;1-z\right) .
	\end{align}


Unlike the non-KR BTZ black hole, when scalar clouds exist in the rotating KR BTZ black hole, the black hole mass and angular velocity cannot be determined analytically. To determine the existence lines of scalar clouds in the KR BTZ black hole, we employ the following numerical procedure. We begin with the KR parameter $\ell = 0$, reducing the system to the non-KR case, and employ the background mass $M(\Omega_H=0)$ for scalar clouds as initial seed. In the non-KR BTZ black hole, scalar clouds exist only when the mixing parameter $\xi$ of the Robin boundary~\eqref{eq:robin} satisfies the following relation~\cite{Ferreira:2017cta}
	\begin{equation}
		tan\left (\xi\right)=\frac{\Gamma \left (\sqrt{1+\mu ^{2}} \right ) \left |\Gamma \left (\frac{1}{2}-\frac{1}{2}\sqrt{1+\mu ^{2}}+i \frac{m \lambda}{2\sqrt{r_{+}}}\right)\right|^{2}}{\Gamma \left (-\sqrt{1+\mu ^{2}} \right ) \left |\Gamma \left (\frac{1}{2}+\frac{1}{2}\sqrt{1+\mu ^{2}}+i \frac{m \lambda}{2\sqrt{r_{+}}}\right)\right|^{2}} .
	\end{equation}
In this case, the black hole mass has a simple analytical expression
\begin{equation}
	M=\frac{r_{+}}{\lambda ^{2}} \left ( 1+\lambda ^{2} \Omega_H^{2} \right ) .
\end{equation}
We then gradually increase the KR parameter $\ell$, while keeping the horizon angular velocity fixed at $\Omega_H=0$, and at each step use the solution obtained at the previous value of $\ell$ as the initial guess to numerically solve the Wronskian determinant~\eqref{eq:wronskian_def}. Finally, for each fixed value of $\ell$, we increase the angular velocity $\Omega_H$ incrementally starting from the solutions obtained above, thereby obtaining the complete existence lines and determining the corresponding background mass $M$. For concreteness, in the following we will set $\mu^{2}=-0.65$. As a matter of fact,  other choices will not qualitatively modify our results.

\begin{figure}[htbp]
	\centering
	\begin{minipage}[b]{0.32\textwidth}
		\centering
		\includegraphics[width=\textwidth]{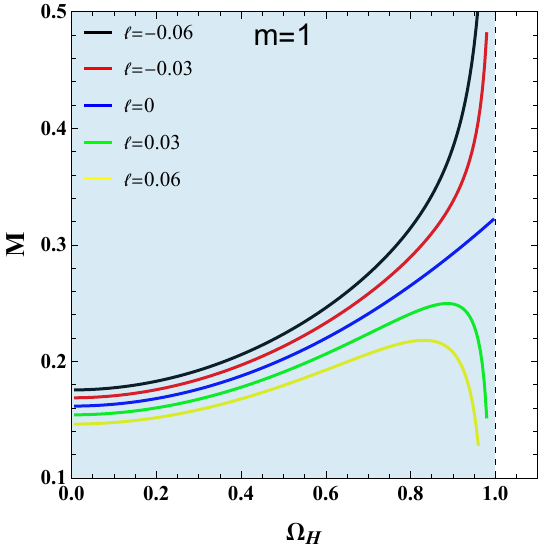}
	\end{minipage}
	\begin{minipage}[b]{0.32\textwidth}
		\centering
		\includegraphics[width=\textwidth]{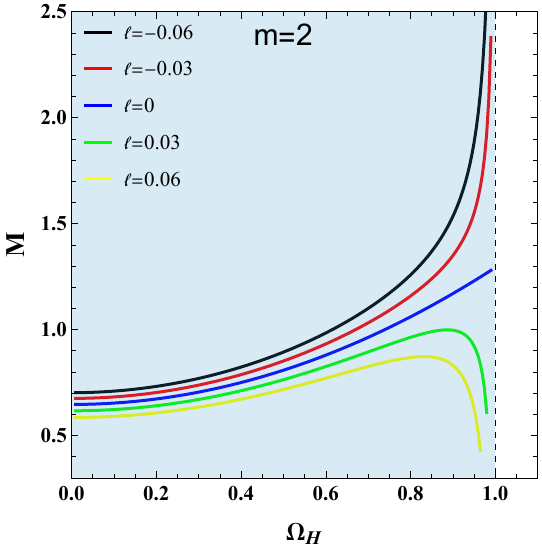}
	\end{minipage}
	\begin{minipage}[b]{0.315\textwidth}
		\centering
		\includegraphics[width=\textwidth]{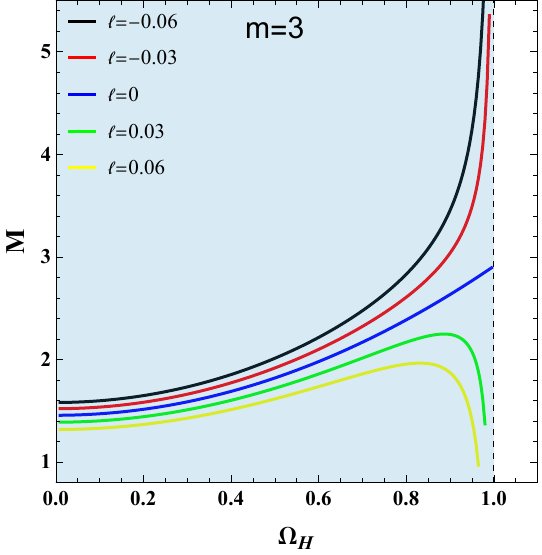}
	\end{minipage}	
	\caption{Existence lines of nodeless scalar clouds ($n=0$) for various KR parameters $\ell$ with fixed quantum numbers $m=1$, $2$ and $3$ in the mass vs horizon angular velocity parameter space of KR BTZ black holes. Here we set the AdS radius $\lambda=1$. The dashed black curve corresponds to extremal KR BTZ black holes with $\Omega_H =1/\lambda$ and non-extremal KR BTZ black holes exist in the shaded region.
	}
	\label{fig:cloud1}
\end{figure}

\begin{figure}[htbp]
	\centering
	\begin{minipage}[b]{0.32\textwidth}
		\centering
		\includegraphics[width=\textwidth]{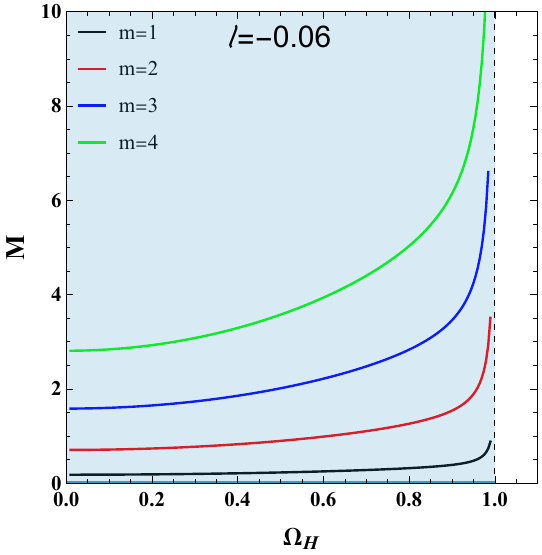}
	\end{minipage}
	\begin{minipage}[b]{0.316\textwidth}
		\centering
		\includegraphics[width=\textwidth]{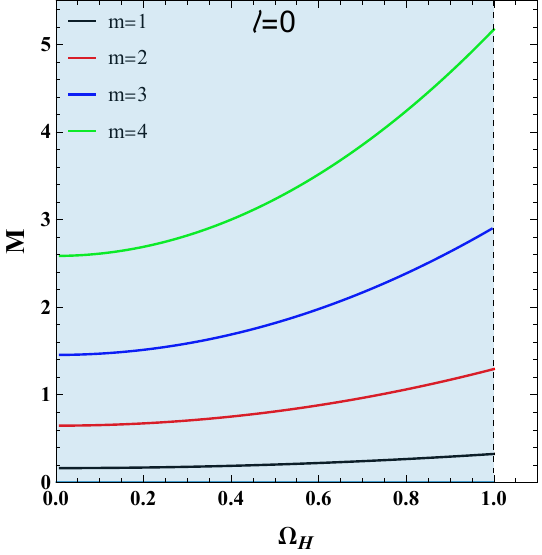}
	\end{minipage}		
	\begin{minipage}[b]{0.323\textwidth}
		\centering
		\includegraphics[width=\textwidth]{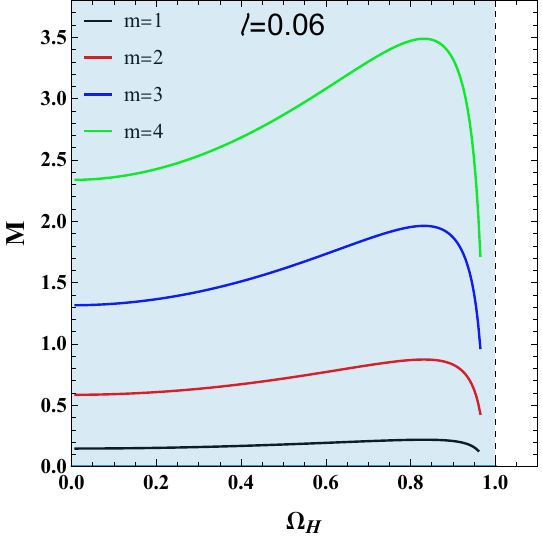}
	\end{minipage}		
	\caption{Existence lines of nodeless scalar clouds ($n=0$) for various quantum numbers $m$ with fixed KR parameters $\ell=-0.06$, $0$ and $0.06$ in the mass vs horizon angular velocity parameter space of KR BTZ black holes. Here we set the AdS radius $\lambda=1$. The dashed black curve corresponds to extremal KR BTZ black holes with $\Omega_H =1/\lambda$ and non-extremal KR BTZ black holes exist in the shaded region.	
	}
	\label{fig:cloud2}
\end{figure}

To investigate the influence of the KR parameter $\ell$ on scalar clouds, we show in Fig.~\ref{fig:cloud1} several representative cloud configurations for different values of $\ell$ and quantum numbers
$m$, with $\xi=0.9\,\pi$ and $\lambda=1$. The results demonstrate that for a fixed $m$, increasing $\ell$ allows clouds to exist at a smaller black hole mass $M$ for a given angular velocity $\Omega_H$. In addition, we find an interesting phenomenon: when $\ell \leq 0$, for a fixed $M$, the black hole can support a stationary scalar cloud only at a single $\Omega_H$, regardless of the value of $m$. By contrast, when $\ell > 0$, for the same fixed $M$, scalar clouds may exist at two distinct angular velocities. This behavior suggests that, for $\ell>0$, the KR correction qualitatively modifies the eigenvalue structure of the scalar cloud condition. As a result, for a fixed  mass $M$, the existence line in the $(M,\Omega_{H})$ plane may become non-monotonic, allowing two distinct values of $\Omega_{ H}$ to support scalar clouds. Physically, this indicates the emergence of two different rotational branches of marginally bound configurations. We next investigate the dependence on the quantum number $m$ in Fig.~\ref{fig:cloud2}. In each panel, $\ell$ is fixed while $m$ is varied. For the fixed  angular velocity $\Omega_H$, larger values of $m$ correspond to larger background black hole masses, and this behavior remains unchanged in the presence of $\ell$. 

\begin{figure}[htbp]
	\centering
	\begin{minipage}[b]{0.32\textwidth}
		\centering
		\includegraphics[width=\textwidth]{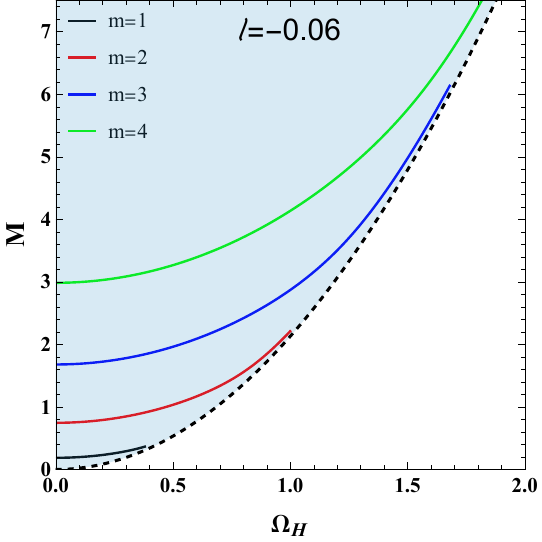}
	\end{minipage}
	\begin{minipage}[b]{0.317\textwidth}
		\centering
		\includegraphics[width=\textwidth]{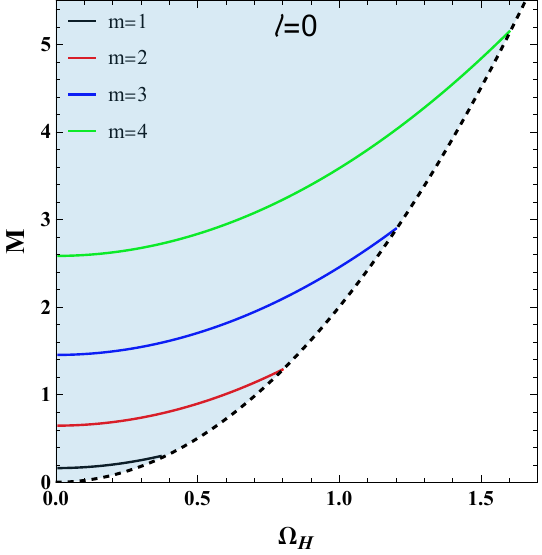}
	\end{minipage}		
	\begin{minipage}[b]{0.323\textwidth}
		\centering
		\includegraphics[width=\textwidth]{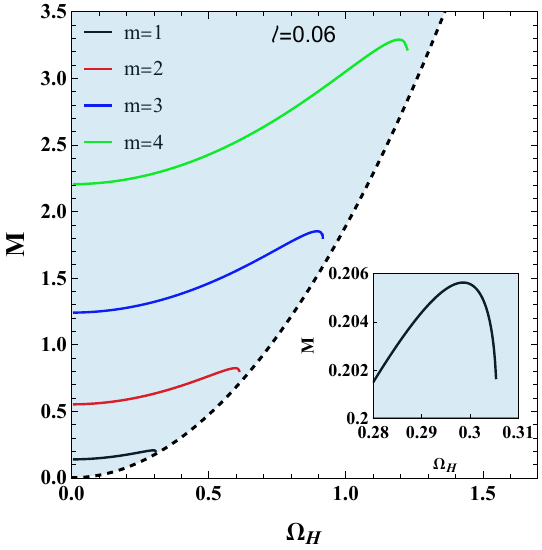}
	\end{minipage}		
	\caption{Existence lines of nodeless scalar clouds ($n=0$) for various quantum numbers $m$ with fixed KR parameters $\ell=-0.06$, $0$ and $0.06$ in the mass vs horizon angular velocity parameter space of KR BTZ black holes.  Here we set the horizon radius $r_{+}=1$. The dashed black curve corresponds to extremal KR BTZ black holes with $\Omega_H=1/\lambda$ and non-extremal KR BTZ black holes exist in the shaded region.	
	}
	\label{fig:cloud3}
\end{figure}

{We also fix the horizon radius $r_+$, rather than the AdS radius $\lambda$, to examine how the KR parameter $\ell$ and the quantum number $m$ affect the scalar clouds.} In Fig.~\ref{fig:cloud3}, we show existence lines with $\xi=0.9\,\pi$ and $r_+=1$. The results confirm a similar trend: for a fixed $\ell$,  if increasing the quantum number $m$, the existence lines shift toward larger $M$ for a given angular velocity $\Omega_H$. The same qualitative behavior is also observed with a fixed $r_{+}$: for $\ell\leq0$, a given mass $M$ supports a stationary scalar cloud at only one value of $\Omega_H$, whereas for $\ell>0$, two distinct values of $\Omega_H$ may exist for the same $M$.
The results presented in {the} above figures indicate that {both} the KR parameter $\ell$ and the quantum number $m$  affect the existence lines of nodeless scalar clouds $(n=0)$, this inverse relationship is consistently observed, regardless of whether the background is fixed by $\lambda=1$ or $r_{+}=1$. It should be {noted} that the influence of the KR parameter $\ell$ on the extremal lines (the dashed black curves) depends on our choice: it has no effect when we choose $\lambda=1$, while distinctly shifting the position of the extremal line when we choose $r_{+}=1$.
    
\begin{figure}[htbp]
	\centering
	\begin{minipage}[b]{0.45\textwidth}
		\centering
		\includegraphics[width=\textwidth]{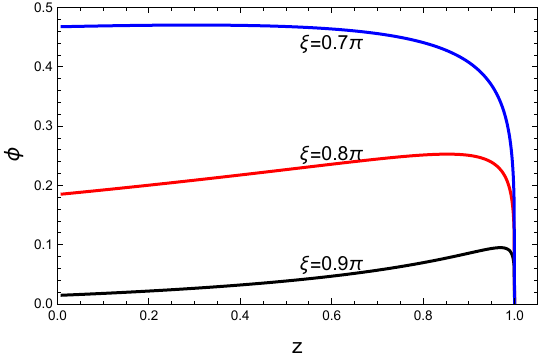}
	\end{minipage}
	\begin{minipage}[b]{0.455\textwidth}
		\centering
		\includegraphics[width=\textwidth]{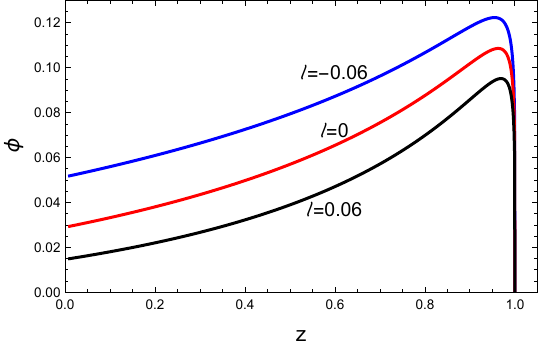}
	\end{minipage}		
	\caption{Radial profiles of degenerate clouds on KR BTZ black holes are presented in terms of $z$ with $m=1$, $\lambda=1$, $r_+=0.114776$ and $r_-=0.0929685$. The left panel is for different Robin boundary conditions $\xi= 0.7\pi$, $ 0.8\pi$ and $ 0.9\pi$ at infinity with $\ell=0.06$, and the right panel is for different KR parameters $\ell=-0.06, 0$ and $0.06$ with $\xi= 0.9\pi$.
	}
	\label{RadialProfile}
\end{figure}

In Fig.~\ref{RadialProfile}, we display representative radial profiles of stationary scalar clouds for selected parameters. These profiles correspond to solutions on the existence lines of scalar clouds. One can see more specifically that, for the fixed $\mu^2$ and $m$, increasing the Robin parameter $\xi$ suppresses the overall amplitude of the scalar cloud and makes the profile more strongly peaked before it vanishes at the boundary. By contrast, varying the KR parameter $\ell$ mainly changes the overall magnitude of the profile: the cloud amplitude decreases as $\ell$ increases, while the general shape of the radial profile remains largely unchanged.

So far, our analysis has been restricted to the ground state stationary clouds since we only find the nodeless scalar clouds ($n=0$) when we scan the parameter space of the system for given values
of $(\lambda, \ell, m, \xi)$ or  $(r_{+}, \ell, m, \xi)$. Considering that such clouds exist only at the superradiant threshold, it is natural to turn next to the QNM spectrum of the KR BTZ black hole, in order to clarify the origin of these stationary bound states.

QNM
\subsection{Quasinormal modes}
Similar to the case of scalar clouds, the analytical QNM spectrum is not available for the KR BTZ spacetime. Nevertheless, the numerical approach can also be employed. The numerical treatment of the QNMs closely parallels that used for the scalar clouds and will not be repeated here. Starting from the known QNM frequencies at $\ell=0$ under Dirichlet or Neumann boundary conditions, we continuously deform the solutions to $\ell\neq0$ and then further extend them to general Robin boundary conditions by varying $\xi$, thereby obtaining the QNM spectrum $\omega_n(\ell,\xi)$. To set the initial value for the numerical analysis, we first consider the $\ell=0$ limit, in which the Heun function reduces to the Gauss hypergeometric function ${}_2F_1$. In this case, the QNMs of the non-KR BTZ black hole under Dirichlet and Neumann boundary conditions can be obtained analytically~\cite{Dappiaggi:2017pbe}

\begin{align}
	\omega_{n}^{(\mathcal{D}),\mathrm{L}} &= \frac{m}{\lambda} - i \frac{(\sqrt{r_{+} }- \sqrt{r_{-}})}{\lambda^2} \left(2n + 1 + \sqrt{1 + \mu^2}\right), \label{eq:QNMDL} \\
	\omega_{n}^{(\mathcal{D}),\mathrm{R}} &= -\frac{m}{\lambda} - i \frac{(\sqrt{r_{+} }+ \sqrt{r_{-}})}{\lambda^2} \left(2n + 1 + \sqrt{1 + \mu^2}\right), \label{eq:QNMDR}\\
	\omega_{n}^{(\mathcal{N}),\mathrm{L}} &= \frac{m}{\lambda} - i \frac{(\sqrt{r_{+} }- \sqrt{r_{-}})}{\lambda^2} \left(2n + 1 - \sqrt{1 + \mu^2}\right), \label{eq:QNMNL} \\
	\omega_{n}^{(\mathcal{N}),\mathrm{R}} &= -\frac{m}{\lambda} - i \frac{(\sqrt{r_{+} }+ \sqrt{r_{-}})}{\lambda^2} \left(2n + 1 - \sqrt{1 + \mu^2}\right), \label{eq:QNMNR}
\end{align}
where the labels $\mathrm{L}$ and $\mathrm{R}$ identify the left and right branches, respectively.

\begin{figure}[tbp]
	\centering
	\begin{minipage}[b]{0.32\textwidth}
		\centering
		\includegraphics[width=\textwidth]{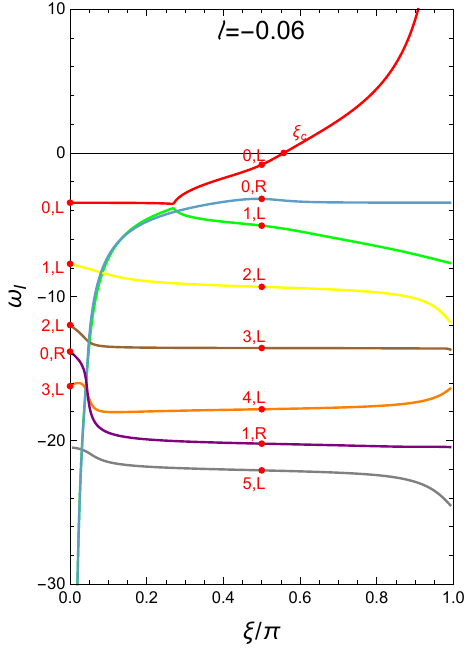}
	\end{minipage}
	\begin{minipage}[b]{0.32\textwidth}
		\centering
		\includegraphics[width=\textwidth]{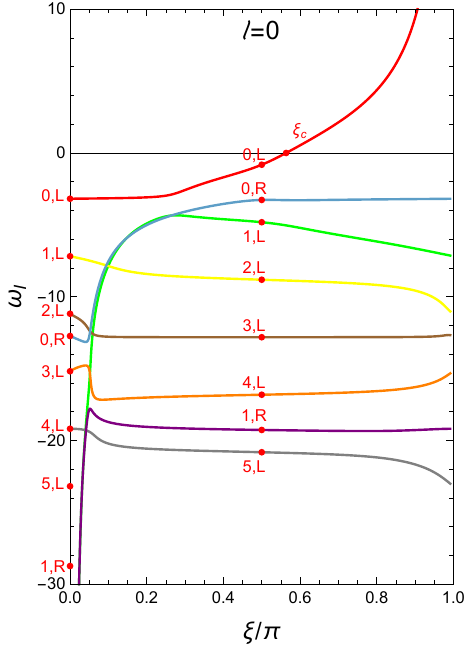}
	\end{minipage}
	\begin{minipage}[b]{0.32\textwidth}
		\centering
		\includegraphics[width=\textwidth]{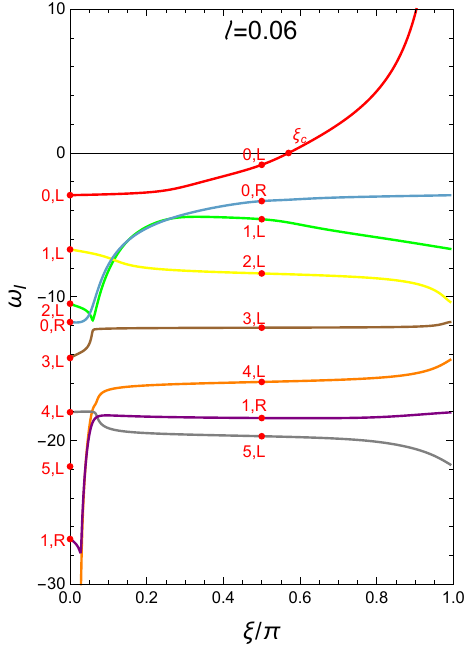}
	\end{minipage}
	\begin{minipage}[b]{0.32\textwidth}
		\centering
		\includegraphics[width=\textwidth]{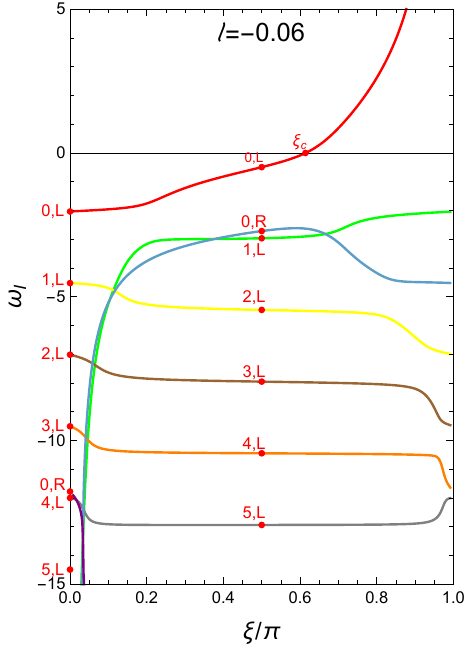}
	\end{minipage}
	\begin{minipage}[b]{0.32\textwidth}
		\centering
		\includegraphics[width=\textwidth]{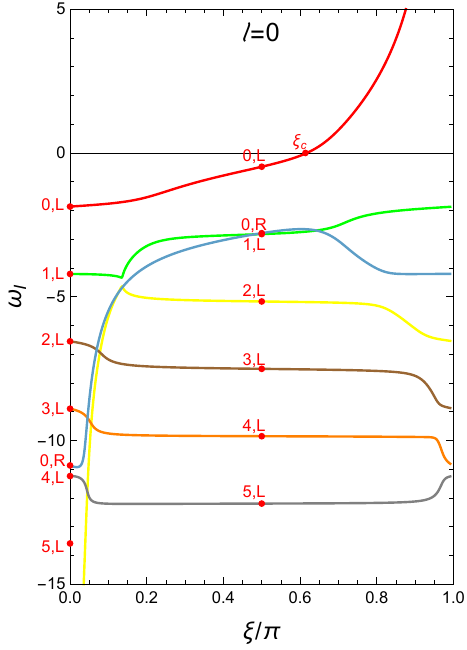}
	\end{minipage}
	\begin{minipage}[b]{0.32\textwidth}
		\centering
		\includegraphics[width=\textwidth]{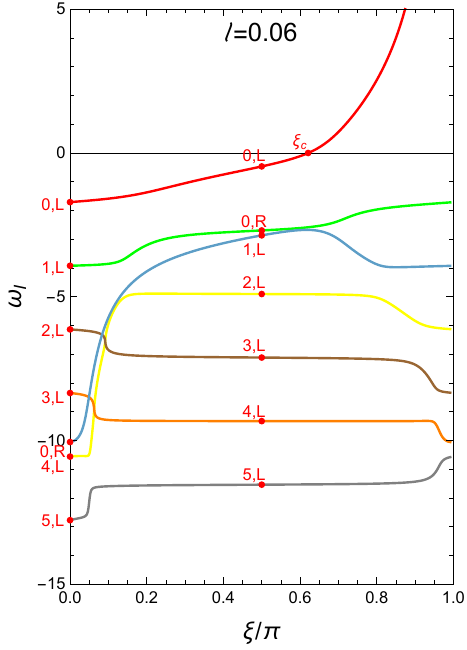}
	\end{minipage}
	\caption{Imaginary parts of some QNM frequencies as a function of $\xi/\pi$ for a KR BTZ black hole with different KR parameters $\ell=-0.06$, $0$, $0.06$ with $m=1$. The upper panels correspond to the case with $\lambda=1$, $r_{+}=25$, $r_{-}=9$ where the threshold values $\xi_{c}/\pi$ are $0.557$, $0.564$ and $0.570$ as $\ell$ changes from $-0.06$ to $0.06$, and the lower panels to $r_{+}=1$, $r_{-}=0.5$, $\lambda=0.5$ where the threshold values  $\xi_{c}/\pi$ are $0.614$, $0.618$ and $0.621$, respectively. }
	\label{fig:QNM}	
\end{figure}

The effect of the KR parameter $\ell$ on the QNM spectrum has already been discussed in previous work~\cite{Xia:2025yzg}, so we do not repeat a detailed analysis here. Instead, our main interest is the condition for scalar clouds, namely the onset point at which $\omega_I=0$. In Fig.~\ref{fig:QNM}, we present several representative QNM {spectra}. By scanning a broad range of parameter space, we find a robust and qualitatively similar pattern: as illustrated in the figure, only the fundamental nodeless mode ($n=0$) can develop an instability when the Robin mixing parameter $\xi$ exceeds a critical value $\xi_c$, i.e., when $\xi>\xi_c$ one has $\omega_I>0$. The threshold value $\xi_c$ separating the stable and unstable regimes is precisely where the stationary scalar clouds exist. The figure also shows that, when $\ell=-0.06$, $0$ and $0.06$, the threshold values are $\xi_{c}/\pi=0.557$, $0.564$ and $0.570$ in the case of $\lambda=1$, but $\xi_{c}/\pi=0.614$, $0.618$ and $0.621$ in the case of $r_{+}=1$. Thus, we conclude that, with the other parameters fixed, the threshold value $\xi_c$ increases with the increase of the KR parameter $\ell$, regardless of $\lambda=1$ or $r_{+}=1$, which means that varying the KR parameter $\ell$ significantly shifts the critical value $\xi_c$. This explains why $\ell$ has such a pronounced effect on the existence lines and radial profiles of scalar clouds in the previous subsection, and also justifies our focus there on the nodeless mode ($n=0$)  only.

\subsection{Superradiant threshold}

To better understand the physical origin of the stationary scalar clouds constructed above, we now briefly examine the horizon energy and angular-momentum fluxes. This allows us to relate the cloud solutions to the corresponding threshold behavior of the scalar perturbations. It is convenient to employ ingoing Eddington--Finkelstein (EF) coordinates $(v,r,\hat{\varphi})$.  The coordinate transformation is
	\begin{equation}
		dv = dt + dr_* \doteq dt + \frac{dr}{A},\qquad
		d\hat{\varphi} = d\varphi - \frac{K}{A}\,dr,
	\end{equation}
with the tortoise coordinate
	\begin{equation}
		r_* = \frac{1+\ell}{2\left(r_+^2-r_-^2\right)}
		\Bigg[r_+ \log\!\Big(\frac{r-r_+}{r+r_+}\Big)
		- r_- \log\!\Big(\frac{r-r_-}{r+r_-}\Big)\Bigg].
	\end{equation}
So we have  $r_* \to -\infty$ as $r\to r_+$, which ensures that EF time slices can extend smoothly through the horizon. In EF coordinate, we employ an ingoing wave boundary condition
at the event horizon and rewrite the scalar field solution as
	\begin{equation}
		\Phi(v,r_+,\hat{\varphi})=\phi^{(I)}(z)\,e^{-i\omega t + i m\varphi}
		= \Bigg(\frac{4r_+^2}{r_+^2-r_-^2}\Bigg)^{\!\alpha}\,e^{-i\omega v + i m \hat{\varphi}},
	\end{equation}
where the exponent $\alpha$ can be fixed by the boundary condition. Since the energy momentum tensor of the massive scalar field is
	\begin{equation}
		T_{\mu\nu}
		= \partial_{(\mu}\bar{\Phi}\,\partial_{\nu)}\Phi
		- \tfrac{1}{2}\,g_{\mu\nu}\!\left[g^{\rho\tau}\partial_{(\rho}\bar{\Phi}\,\partial_{\tau)}\Phi
		+ \tfrac{1}{2}\frac{\mu^2}{\lambda^2}|\Phi|^2\right],
	\end{equation}
we can obtain the energy flux at the horizon
	\begin{equation}
		\mathcal{F}_E(v)
		=\int_0^{2\pi} d\hat{\varphi}\; r_+\,T_{\mu\nu}\,\chi^\mu k^\nu
		= F\big[\omega_I^2+\omega_R(\omega_R-m\Omega_H)\big]\,e^{2\omega_{I} v},\label{eq:FE}
	\end{equation}
with 
	\begin{equation}
	F = 2\pi r_+ \Bigg(\frac{4r_+^2}{r_+^2-r_-^2}\Bigg)^{2\alpha},
\end{equation}
where  $k^\nu = \partial^\nu_r$ is the radial basis vector and $\chi^\mu = \partial_t + \Omega_H\partial_{\hat\varphi}$ is the horizon Killing vector. The energy flux across the horizon, given in~\eqref{eq:FE}, can become negative when $
	\omega_R(m\Omega_H-\omega_R)>\omega_I^2$, signifying that the energy is coming out of the black hole towards the exterior region, which is superradiance. Similarly, the flux of angular momentum at the horizon is
	\begin{equation}
		\mathcal{F}_L(v)
		= -\int_0^{2\pi} d\hat{\varphi}\; r_+\,T_{\mu\nu}\,\chi^\mu m^\nu
		= F\,m(\omega_R-m\Omega_H)\,e^{2\omega_I v},
	\end{equation}
where $m^\nu=\partial^\nu_{\hat\varphi}$ is the axial Killing vector. Obviously, the angular momentum is always towards the exterior region when $\omega_R<m\Omega_H$. Analyzing both the signs and temporal evolution of $\mathcal{F}_E$ and $\mathcal{F}_L$ enables an unambiguous distinction between the superradiant extraction and bulk instabilities, which provides a foundation for the stability criterion of the KR BTZ spacetime.

    \begin{figure}[htbp]
		\centering
		\begin{minipage}[b]{0.475\textwidth}
			\centering
			\includegraphics[width=\textwidth]{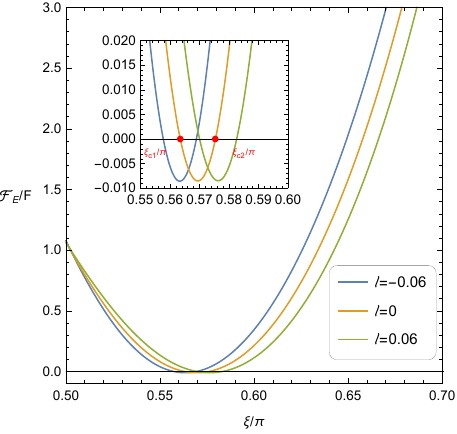}
		\end{minipage}
		\begin{minipage}[b]{0.48\textwidth}
			\centering
			\includegraphics[width=\textwidth]{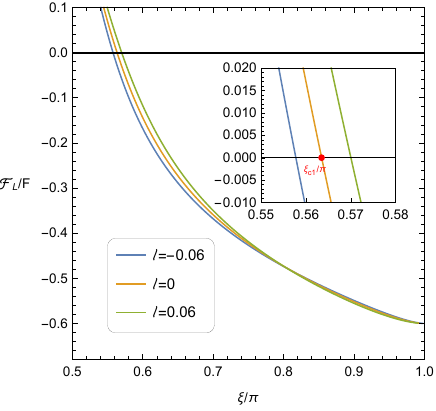}
		\end{minipage}
		\begin{minipage}[b]{0.488\textwidth}
			\centering
			\includegraphics[width=\textwidth]{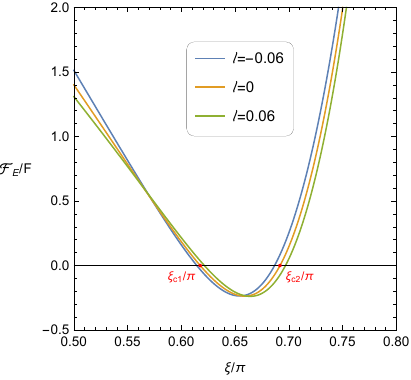}
		\end{minipage}
		\begin{minipage}[b]{0.485\textwidth}
			\centering
			\includegraphics[width=\textwidth]{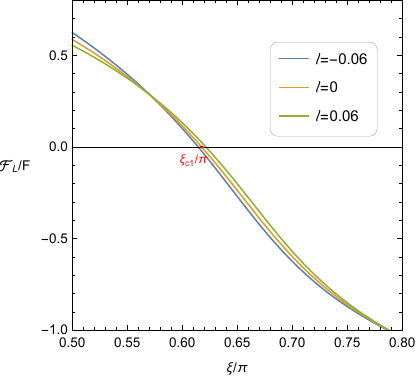}
		\end{minipage}		
	\caption{Energy and angular momentum fluxes as a function of $\xi/\pi$ for a KR BTZ black hole with different KR parameters $\ell=-0.06$, $0$, $0.06$ with $n=0$ and $m=1$. The upper panels correspond to the case with $\lambda=1$, $r_{+}=25$, $r_{-}=9$ where the threshold values $\xi_{c1}/\pi=0.564$ and $\xi_{c2}/\pi=0.575$ for $\ell=0$, and the lower panels to $r_{+}=1$, $r_{-}=0.5$, $\lambda=0.5$ where the threshold values $\xi_{c1}/\pi=0.618$ and $\xi_{c2}/\pi=0.692$ for $\ell=0$. }
		\label{fig:sup}	
	\end{figure}
	
In Fig.~\ref{fig:sup}, we show the horizon energy and angular momentum fluxes, $\mathcal{F}_E$ and $\mathcal{F}_L$, as functions of $\xi/\pi$ for three KR deformations, $\ell=-0.06$, $0$, and $0.06$, with $\lambda=1$ (upper panels) and $r_{+}=1$ (lower panels). This figure is particularly useful for clarifying the physical origin of the scalar clouds. In asymptotically AdS spacetimes, a mode with $\omega_I>0$ does not necessarily correspond to  superradiance, since it may also be related to a bulk AdS instability. By contrast, the scalar clouds should appear precisely at the threshold $\xi_{c1}$ (as presented in Fig.~\ref{fig:sup}, for example, $\xi_{c1}/\pi=0.564$ when $\ell=0$ in the case of $\lambda=1$, but $\xi_{c1}/\pi=0.618$ when $\ell=0$ in the case of $r_{+}=1$), where the black hole starts to transfer the energy and angular momentum to the scalar field. As shown by comparison with Fig.~\ref{fig:QNM}, when $\xi$ reaches the same critical value $\xi_c=\xi_{c1}$, the QNM frequency satisfies $\omega_I=0$, while both $\mathcal{F}_E$ and $\mathcal{F}_L$ simultaneously reverse sign. This identifies $\xi_{c1}$  as the superradiant threshold and hence as the location of the stationary scalar clouds found in the previous section. For larger $\xi$, $\omega_I$ may stay positive even after the energy flux becomes positive again when $\xi>\xi_{c2}$ (as shown in Fig.~\ref{fig:sup}, for example, $\xi_{c2}/\pi=0.575$ when $\ell=0$ in the case of $\lambda=1$, but $\xi_{c2}/\pi=0.692$ when $\ell=0$ in the case of $r_{+}=1$), indicating that this unstable region is not directly associated with the scalar clouds but instead with a bulk AdS instability. The KR parameter $\ell$ does not alter the basic physical picture, but it changes the superradiant critical values and intervals quantitatively. In particular, regardless of $\lambda=1$ or $r_{+}=1$, varying $\ell$ shifts the cloud threshold $\xi_{c1}$ and modifies the range of the superradiant regime ($\xi_{c1}$, $\xi_{c2}$).

\section{Conclusion}

In this work, we have investigated massive scalar perturbations in the rotating KR BTZ black hole background, with particular emphasis on stationary scalar clouds, QNMs and superradiant instabilities. By separating variables and performing a suitable radial transformation, we reduced the Klein--Gordon equation to a general Heun equation. Imposing ingoing behavior at the horizon together with vanishing energy flux boundary conditions at infinity, we determined the existence lines of stationary scalar clouds and the corresponding QNM spectrum through the Wronskian determinant. Our analysis shows that the KR BTZ black hole can indeed support stationary fundamental scalar clouds under general Robin boundary conditions, and that the KR parameter $\ell$ has a substantial impact on the existence lines and radial profiles of scalar clouds. In particular, the KR parameter $\ell$ and the azimuthal quantum number $m$ affect the cloud existence lines in opposite directions, while the sign of $\ell$ can also lead to qualitatively different patterns in the parameter space, i.e., regardless of fixing the AdS radius $\lambda=1$ or the horizon radius $r_{+}=1$, for $\ell\leq0$, a given mass $M$ supports a stationary scalar cloud at only one value of the horizon angular velocity $\Omega_H$, whereas for $\ell>0$, two distinct values of $\Omega_H$ may exist for the same $M$. In the limit $\ell\to0$, our results reduce to those of the rotating BTZ black hole.

To clarify the dynamical origin of these cloud configurations, we further analyzed the QNM spectrum and the associated horizon fluxes. We found that only the fundamental mode on the left branch can develop an instability with a critical Robin parameter $\xi_c$, and that this threshold is significantly shifted by the non-zero KR parameter. More importantly, the stationary scalar clouds are located precisely at the onset of superradiance, where the imaginary part of the frequency vanishes and the horizon energy and angular momentum fluxes simultaneously change sign. This confirms that the cloud solutions correspond to the superradiant threshold. At the same time, our flux analysis shows that not every unstable mode in asymptotically AdS spacetime is superradiant: in part of the unstable region, the mode growth is instead associated with the AdS bulk instability rather than with the energy extraction from the black hole. Overall, these results show that the KR parameter $\ell$ plays the role of an effective control parameter for the formation of scalar clouds, the onset of superradiance, and the perturbative stability of the KR BTZ black hole.

There are several natural directions for  future work. Since scalar clouds represent the threshold configurations of superradiance, an important next step is to go beyond the test  field  and include the backreaction of sufficiently amplified scalar fields on the geometry, in order to explore whether the KR BTZ black hole can support genuinely hairy solutions and how the KR parameter $\ell$ modifies their existence and properties. In addition, it would also be interesting to extend the present analysis to spin-1 fields, such as massive vector (Proca) perturbations, and examine whether vector clouds or vector hair may arise in the KR background. Such extensions may reveal new features of the cloud formation.

\begin{acknowledgments}		
	
	This work was supported by the National Natural Science Foundation of China (Grant Nos. 12275079 and 12035005), the National Key Research and Development Program of China (Grant No. 2020YFC2201400) and the innovative research group of Hunan Province (Grant No. 2024JJ1006).
	
\end{acknowledgments}

\bibliography{reference2}
\end{document}